\documentclass[usletter,12pt]{article}
\usepackage{verbatim}

\usepackage{graphicx}
\usepackage{amsmath}
\usepackage{epsfig}
\usepackage{a4wide}
\usepackage{amsfonts}
\usepackage{amssymb}
\usepackage{epsfig}
\def\stu{St\"uckelberg }

\def\ie{{\it i.e.}}

\def\ba{\begin{eqnarray}}
\def\ea{\end{eqnarray}}
\def\mpl{M_{\rm Pl}}

\def\d{\mathrm{d}}
\def\p{{\cal P}}

\def\K{{\cal K}}

\def\L*{{\cal L}_*}
\def\L{\mathcal{L}}
\def\({\left(}
\def\){\right)}
\def\ie{{\it i.e. }}
\def\eg{{\it e.g. }}

\def\nn{\nonumber}
\def\p{\partial}
\def\mn{_{\mu \nu}}
\def\stu{St\"uckelberg }
\def\p{\partial}
\def\mupn{^\mu_{\ \nu}}
\def\<{\langle}
\def\>{\rangle}

\begin{document}
\title{{\Large \bf Galileons in the Sky}\footnote{Accepted review article to appear in a special volume of the ``Comptes
Rendus de l'Academie des Sciences" about Dark Energy and Dark Matter.}\ \footnote{Based on work in collaboration with C.~Burrage, G.~Gabadadze, L.~Heisenberg, D.~Pirtskhalava, D.~Seery, A.J.~Tolley and I.~Yavin.} }
\author{\bf Claudia de Rham\\[5pt]
D\'epartment de Physique  Th\'eorique and Center for
Astroparticle Physics, \\
Universit\'e de  Gen\`eve, 24 Quai E. Ansermet, CH-1211  Gen\`eve, Switzerland \\
\& \\
Department of Physics, Case Western Reserve University,\\
Euclid Ave,
Cleveland, OH, 44106, USA}

\date{}

\maketitle

\abstract{We review the different frameworks in which Galileon scalar fields have been seen to emerge such an in DGP, New Massive Gravity and Ghost-free massive Gravity and emphasize their relation with the Lovelock invariant in braneworld models. The existence of a non-renormalization theorem for Galileon scalar fields makes them especially attractive candidates for inflation as well as for late-time acceleration. In particular we review the self-accelerating and degravitating branches of solutions present in Galileon models when arising from Massive Gravity and discuss their phenomenology.}

\section{Introduction}

The late-time acceleration of the Universe remains undeniably one of the most intriguing aspects of modern Cosmology, at the interface between General Relativity and  particle physics. Over the past decade a multitude of new models have been developed
all across the community to tackle this fascinating issue, see
Refs.~\cite{Carroll:2000fy,Copeland:2006wr,Durrer:2007re} for different reviews as well as Refs.~\cite{Astier:Pain,Kunz,Martin,Clarkson} for other invited contributions to the  CRAS special issue on Dark matter/Dark energy.

Despite the abundance of new scenarios, one can only really distinguish three possibilities: Either the acceleration of the Universe is driven by a Cosmological Constant,
or dark energy is dynamical, in which case the acceleration of the Universe relies on the existence of new degrees of freedom (dof),
or finally, the acceleration of the Universe relies on dofs already accounted for in Cosmology (\eg from the backreaction of inhomogeneities).
Furthermore, independently of whether or not dark energy is dynamical,  new degrees of freedom are usually required to tackle the Cosmological Constant problem and explain why the vacuum energy is either so much smaller than theoretically anticipated or why its gravitational effect is so weak.

Independently of how these new degrees of freedom tackle the Cosmological Constant problem, the late-time acceleration of the Universe or possibly the Coincidence problem, their mere presence raises a new puzzle of its own: How could they possibly affect physics on cosmological scales and yet remain hidden on smaller distance scales. From energy scales of about $10^{-13}$eV all the way up to TeV scales, the presence of an additional scalar field which would mediate a fifth force is extremely well constrained, be it from solar system tests, laboratory tests of General Relativity, or cosmological tests such as Big Bang Nucleosynthesis, to name only a few.

The range of a field is given by its Compton wavelength, thus a scalar field can only be relevant on today's cosmological scales if its mass is of the order or smaller than the Hubble parameter today. With such a small mass, it is surprising that such a degree of freedom would not have already been detected. Over the years, several frameworks have been developed to explain how and why light scalar fields could be relevant on cosmological scales and yet not manifest themselves on solar system scales and every day experiments.
The most direct way to hide such a scalar is simply to prevent any coupling between that field and the standard model. However coupling to gravity usually makes this framework radiatively unstable unless the field is protected by a symmetry, such as for instance the discrete shift symmetry present in pseudo-Goldstone bosons, which could provide a radiatively stable realization of quintessence.

If, on the other hand, the scalar field couples to matter, the field should be hidden via a so-called ``screening" mechanism. One of the first realization of such a mechanism was proposed by Damour and Polyakov, \cite{Damour:1994zq} within the ``Least Coupling Principle", providing a relaxation mechanism where the dilaton effectively decouples from matter. Besides the Least Coupling Principle, only three screening mechanisms are known:
The Chameleon \cite{Khoury:2003aq}, the Symmetron \cite{arXiv:0709.3825,Hinterbichler:2010es} and finally the Vainshtein mechanism, \cite{Vainshtein:1972sx}, see Ref.~\cite{Khoury:2010xi} for a review.
The underlying physics and screening behind these three different mechanisms is very distinct and we shall summarize them briefly in this review, however they all have in common the characteristic of changing behavior depending on the medium (or on the energy scale of the medium).  In the Chameleon, for instance, its effective mass depends on the energy density of the environment. In the Symmetron and Vainshtein mechanisms, on the other hand, it is their effective coupling to matter (or external sources) which is affected by the medium, although the explicit realization in the Symmetron and in the Vainshtein are very different.

Whilst any of these previous mechanisms may be realized for scalar fields in their own right, the Vainshtein mechanism is also expected to be naturally realized in modified theories of gravity where the graviton propagates more degrees of freedom, and in particular in theories of massive gravity and their variants.  In this review we will first summarize in section \ref{sec:Screening} these three screening mechanisms and then focus on the Vainshtein effect. The Vainshtein mechanism is seen explicitly at work in ``Galileon" theories and the rest if this review will hence be devoted to Galileons, starting by its different realizations in section \ref{sec:Galileons} before moving to its impact on Cosmology in section \ref{sec:cosmology}.

\section{Screening Mechanisms}
\label{sec:Screening}

There is little doubts that new physics will appear beyond the Standard Model, some of which, such as Dark Matter are expected to have significant effects on the Universe. Nevertheless, these new fields are typically expected to be relatively massive and thus of short range.   These new particles are thus usually expected to bear little effects on large cosmological distance scales, of the order of $10^{2-3}$Mpc or greater. When dealing with dark energy, on the other hand, only fields which remain coherent on such large scales  are expected to have an effect, which is the reason why light scalars have been the primary choice. However the strength with which such new degrees of freedom can couple to the rest of the standard model is very tightly constrained by
searches for fifth forces and violations of the weak equivalence principle. These tests put constraints on the strength of the interactions mediated by such fields to be many orders of magnitude weaker than gravity itself.

Rather than tuning the coupling between new fields and the Standard Model to be small, we take here a different approach and explore the possibility for some scalar fields to keep a natural coupling to the Standard Model, whilst still being in agreement with observations thanks to a screening mechanism. As we shall see, even though only three different classes of mechanisms are known to date, their explicit realization can take many facets. However they all share in common the existence of non-linear interactions which are essential for the screening to work.

Starting with a generic scalar field coupled to an external source $T$ (typically the trace of the stress-energy tensor of the external sources), the action for a canonically normalized scalar field $\phi$ at the linearized level is\footnote{We focus on local and Lorentz invariant scenarios. In this review we also focus on single scalar field scenarios for simplicity, although most of these frameworks are easily generalizable to multiple fields.}
\ba
\L=-\frac 12 (\p \phi)^2 -\frac12 m^2 \phi^2 +\frac{\alpha}{\mpl}\phi T\,,
\ea
where $\alpha$ is the dimensionless coupling constant.
Considering for instance a localized source at $r=0$, $T=M \delta^{(3)}(r)$, then this field typically contributes to the gravitational potential with an additional Yukawa-type contribution of the form $V_{\phi}=\frac{\alpha  M}{\mpl^2} \frac{e^{- m r}}{r} $. At this level, the only way this field can mediate a force which is many orders of magnitude smaller than the gravitational one is either to make it very massive (submillimeter tests of Newton's law then impose $m>$TeV) or to make its coupling to matter (unnaturally) small, $\alpha \ll 1$. However we can also imagine a situation where the characteristics of the field are medium-dependent, for instance the effective mass or the effective coupling (or both) can depend on the energy density of the local environment, either locally $m=m(\rho)$, $\alpha=\alpha(\rho)$ as will be the case in the Chameleon or the Symmetron, or through non-local effects as will be the case in the Galileon.

In particular we will be considering a generic scalar field, conformally coupled to matter with potential $V$ and wave function $Z\mn$,
\ba
\label{general_L}
\L=-\frac 12 Z^{\mu\nu}(\phi)\p_\mu \phi \p_\nu \phi-V(\phi)+A(\phi) T\,.
\ea
In the three mechanisms presented below, we will see how different screening mechanisms may be realized by appropriate choices of $Z$, $V$ and $A$, see Ref.~\cite{Andrew}.

\subsection{Chameleon}

The  Chameleon is one of the simplest model that exhibits an explicit screening mechanism, \cite{Khoury:2003aq}. The idea behind the Chameleon is to start with a light field on cosmological scales (in low-energy environments) and increase its (effective) mass in dense regions (such as within the Galaxy) so that in dense regions, the force mediated by such a field is suppressed by a large Yukawa exponential factor. The explicit realization of the Chameleon mechanism relies on self-interactions within the field potential (\ie the field's potential should include a contribution different than the simple mass term $\phi^2$) as well as a conformal coupling with external sources, while the field has a standard kinetic term $Z\mn=g\mn$ (with  $g\mn$ is the space-time metric).

In the presence of non-relativistic matter, we may split the stress-energy into a contribution $\rho$ from the environment, and a contribution $\delta T$ from localized test particles,  $T=-\rho+\delta T$, such that the potential and the conformal coupling in that environment combine to give rise to an effective potential of the form $V_{\rm eff}(\phi)=V(\phi)+A(\phi)\rho$.  In the Chameleon, the functions $A$ and $V$ are chosen such that the effective potential has a stable minimum at $\phi=\phi_0(\rho)$. Then around that minimum,  $\phi=\phi_0+\varphi$, the effective Lagrangian for the scalar field perturbations is of the form
\ba
\L_{\rm Chameleon}^{(\rho)}=-\frac 12 (\p \phi)^2-\frac 12 m^2_{\rm eff}(\rho)\phi^2+\alpha(\rho)\,  \phi \, \delta T\,.
\ea
with the coupling $\alpha=A'(\phi_0(\rho))$ and the effective mass,
\ba
m^2_{\rm eff}(\rho) = V''(\phi_0)+ A''(\phi_0) \rho \,.
\ea
such that the Chameleon becomes more massive in denser environments.

\subsection{Symmetron}

An alternative to the Chameleon mechanism has been proposed recently by K.~Olive and M.~Pospelov in  \cite{arXiv:0709.3825} as well as by K.~Hinterbichler and J.~Khoury in \cite{Hinterbichler:2010es}. This ``Symmetron" mechanism shares common features with the Chameleon, such as a standard kinetic term $Z\mn=g\mn$, a conformal coupling to matter and non-linearities in its potential, but the realization of the screening relies this time on the weakening of the coupling to matter.

As a specific realization proposed in \cite{Hinterbichler:2010es}, consider a quadratic coupling to matter and a standard `Higgs-like' potential,
\ba
V(\phi)&=&-\frac12 \mu^2 \phi^2+\frac 14 \lambda \phi^4\\
A(\phi)&=&1+\frac{\phi^2}{2M^2}\,,
\ea
such that the effective potential seen by the Symmetron depends on the environment
\ba
V_{eff}(\phi)
=\frac{1}{2}\left(\frac{\rho}{M^2}-\mu^2\right)\phi^2+\frac{1}{4}\lambda \phi^4\,,
\ea
where here again $\rho$ is the energy density of the environment. At early times, when the Universe is sufficiently hot, or within dense regions of the Universe, where  $\rho>(\mu M)^2$, the potential has a unique minimum at $\phi=\phi_0=0$, which preserves the symmetry $\phi \to - \phi$ (or its $U(1)$ analogue if the field is complex).
The effective coupling to matter seen by fluctuations $\phi$ around that minimum is given by $\alpha=A'(0)=0$, which explains the weak coupling to external sources. However as the Universe cools down, the local energy density drops below the critical density $\rho < \rho_c=(\mu M)^2$, the extremum $\phi=0$ then becomes a local maximum and the field rolls towards one of the two minima located at $\phi=\phi_0(\rho)$ with $\phi_0^2=(\mu^2-\rho/M^2)/\lambda$ leading to a spontaneous breaking of the $\mathbb{Z}_2$-symmetry. In regions of small density, the effective coupling is then non-negligible $\alpha=A'(\phi_0)=\phi_0/M^2$ and the field can have a relevant effect on its environment.


\subsection{Vainshtein}
\label{sec:Vainshtein}

Finally the last class of screening mechanism that has been investigated recently relies on non-linearities within the kinetic term $Z\mn(\phi)$ rather than within the potential. This Vainshtein mechanism was first investigated within the context of Fierz-Pauli massive gravity, (see refs.~\cite{8638}) as a way to evade the vDVZ discontinuity \cite{vDVZ}, (explaining how the additional helicity excitations decouple in the massless limit of the graviton), \cite{Vainshtein:1972sx}. We will review explicit realizations of the Vainshtein mechanism in what follows, but start here by reviewing its main features. Since the Vainshtein mechanism does not rely on potential interactions, we may simply set the potential $V(\phi)=0$ and consider a standard and natural conformal coupling to matter, $A(\phi)=\phi/\mpl$. The essence of the Vainshtein mechanism then lies in the non-standard kinetic term which is symbolically of the form,
\ba
Z\mn(\phi)\sim g\mn +\frac{1}{\Lambda^3}\p_\mu \p_\nu \phi+\frac{1}{\Lambda^6}(\p_\mu \p_\nu \phi)^2+\cdots
\ea
At low-energy, the higher contributions to the kinetic term are negligible, such that one recovers a standard kinetic term $Z\mn \sim g\mn $ and the field couples to external sources with a Planck scale coupling. At high energy however, when the background configuration for $\phi$ is such that $\p^2 \phi_0 \gg \Lambda^3$, or $Z(\phi_0)\gg 1$,
the higher interactions take over and the main contribution to the kinetic term of the fluctuations $\varphi$ around that configuration are then of the form $Z(\phi_0)(\p \varphi)^2\sim (\p^2 \phi_0/\Lambda^3)^n \(\p \varphi\)^2$, with $n$ an integer depending on the details of setup. The properly canonically normalized field is then symbolically given by $\hat \varphi= \varphi/\sqrt{Z(\phi_0)}$, such that the effective coupling to matter seen by fluctuations $\hat \varphi$ around that configuration is given by $\alpha=(\mpl \sqrt{Z(\phi_0)})^{-1}\ll \mpl^{-1}$. At high energy the field is thus {\it strongly} coupled to itself (its interactions are important), and becomes {\it weakly} coupled to external sources.

The main issue when dealing with higher order kinetic terms is to ensure that no Ostrogradski instabilities arise, \cite{Ostrogradski,Whittaker,astro-ph/0601672}. The first explicit realization of the Vainshtein mechanism free of such an issue was provided in the context of the DGP (Dvali, Gabadadze, Porrati) model \cite{Dvali:2000hr,Deffayet:2001uk}. As we shall see below, the decoupling limit of DGP model is special in that even though higher derivative interactions can be important at a relatively small energy scale $\Lambda \ll \mpl$, this does not lead to any ghost or standard strong coupling pathologies. These features is shared by a more general class of ``Galileon" theories, which has been seen to arise in different models recently.

\section{Galileons}
\label{sec:Galileons}

\subsection{DGP}
\label{DGP}

The first explicit realization of a soft mass theory of gravity was realized within the context of the higher dimensional DGP scenario. Embedded in an infinite  five-dimensional bulk, the DGP model considers a four-dimensional brane with an induced Einstein-Hilbert term. By imposing a hierarchy between the four and five-dimensional Planck scales ($\mpl$ and $M_5$ respectively), $M_5\ll \mpl$, gravity behaves four-dimensional at short distance, up to the cross over scale $r_c=\mpl^2/M_5^3\equiv m^{-1}$, before the five-dimensional effects take over at large distance scales, leading to an effectively massive (soft mass) theory of gravity on the four-dimensional brane.

Since gravity is intrinsically five-dimensional in this scenario, the graviton propagates five degrees of freedom, one of which behaves a scalar from a four-dimensional view point. In the limit where five dimension is ``switched off", \ie $M_5/\mpl \to 0$, one expects to recover four-dimensional standard General Relativity, and this additional scalar degree of freedom should hence decouple.  This is achieved via the non-linear interactions of the extra scalar mode, as originally suggested by Vainshtein.

The realization of this Vainshtein mechanism is best seen within the so-called ``decoupling limit" where the helicity-2 and -0 mode of the graviton decouple (which we denote respectively as $h\mn$ and $\phi$ in what follows). For this we work at low enough energy compared to four-dimensional Planck scale so that the self-interactions of the helicity-2 mode may be neglected, but keep a class of non-linearities for the helicity-0 mode alive. This is achieved by taking the limit $\mpl, M_5\to \infty$, while keeping the strong coupling scale $\Lambda=(m^2\mpl)^{1/3}=M_5^2/\mpl$ for the helicity-0 mode fixed. In this limit, the usual helicity-2
mode of gravity can be treated linearly while the interactions for the extra mode $\phi$ remain relevant.
The resulting effective action for that mode is then, \cite{Luty:2003vm}
\ba
 \label{eq.DGP_decoupling}
 \mathcal{L}_\phi=3 \phi \Box \phi -\frac{1}{\Lambda^3}(\partial \phi)^2 \Box \phi+\frac{2}{\mpl}\phi \, T\,,
\ea
leading to the equation of motion,
\ba
3\Box \phi + \frac{1}{\Lambda^3}\((\Box \phi)^2-(\p_\mu \p_\nu \phi)^2\)=-\frac{1}{\mpl}T\,.
\ea
This decoupling limit has several remarkable features:
\begin{enumerate}
\item The effective action is local (contains only a finite number of relevant interactions).
\item Even though the interactions involve higher derivative terms (included in $\Box \phi$), they enter in a specific combination such that the equations of motion remain second order in derivative and the theory is hence free of the Ostrogradski instability.
\item The equations of motion are invariant under shift and Galileon global transformations $\phi \to \phi+c+v_\mu x^\mu$, where $c$ and $v_\mu$ are constant. This symmetry is inherited from five-dimensional Poincar\'e invariance.
\item As expected, the interactions for $\phi$ remain important up to the much lower energy scale $\Lambda \ll \mpl$.
\item This higher interactions rely on ``irrelevant operators" from a standard EFT point of view, and one cannot in general consider them to be large without going beyond the regime of validity of the theory. Here the situation is different as these interactions are protected against quantum corrections. This makes it possible to set the scale $\Lambda$ to be small and to rely on these interactions becoming large without needing to include other classes of interactions. This is the essence of the ``non-renormalization" theorem formulated in Refs.~\cite{Luty:2003vm} and \cite{Nicolis:2004qq}, which we describe further in the next subsection.
\end{enumerate}

To see the Vainshtein mechanism at work, we can focus on  spherically symmetric static sources, \cite{Deffayet:2001uk}. Taking for instance a localized mass $M$, for which $T=-M \delta^{(3)}(r)$,  we can then solve for $\phi$ explicitly, and the force $F_\phi\sim \phi_0'(r)$ mediated by this field is then given by \cite{Nicolis:2004qq}
\ba
\label{eqphi'}
\phi_0'(r)=\frac{3}{4}r \Lambda^3\(-1\pm \sqrt{1+\frac{8 M}{9 \mpl r^3\Lambda^3}}\)\,,
\ea
focusing on the `$+$'- branch, we see that at large distance scales, when $r\gg r_V$, we recover the  standard Newton square law, $\phi_0'=M/(\mpl r^2)$, while at short distances the interaction term dominates $\Box \phi_0 \gg \Lambda^3$ and the force becomes much weaker compared to the standard gravitational one, $\phi_0'(r)\sim (\frac{M}{2\mpl}\frac{\Lambda^3}{r})^{1/2}$ for $r\ll r_V$, (this force vanishes in the massless limit $\Lambda \to 0$). The cross-over scale between both behaviors is given by the ``Vainshtein" or ``strong-coupling" radius $r_V=(M/\mpl)\Lambda^{-1}$.

Beyond providing an explicit realization of the Vainshtein mechanism, the decoupling limit of DGP also encodes most of the information on the cosmological behavior of this model. In particular the existence of a self-accelerating solution can also be seen in this decoupling limit, (corresponding to a the `$-$'-branch in \eqref{eqphi'}).

As mentioned previously, the decoupling limit of DGP has several remarkable features which makes it possible to exhibit the Vainshtein mechanism without the presence of any ghost or strong coupling issues (only a finite number of interactions remain important and the theory remains under control within the Vainshtein radius). This decoupling limit was recently generalized to the most general set of interactions that share the same set of features, and particular corresponds to higher derivative interactions that satisfy the Galileon symmetry without leading to any ghost problem (no more than two derivatives at the level of the equations of motion), hence
called the ``Galileon" interactions,  \cite{Nicolis:2008in}.
Within this ``Galileon" framework, one can consider a total of four possible interactions,
\ba
\label{L2}
\L_2&=&(\p \phi)^2\\
\L_3&=&(\p \phi)^2\Box \phi\\
\L_4&=&(\p\phi)^2 ((\Box \phi)^2-(\p_\mu \p_\nu \phi)^2)\\
\L_5&=&(\p\phi)^2 ((\Box \phi)^3-3\Box \phi (\p_\mu \p_\nu \phi)^2+2 (\p_\mu \p_\nu \phi)^3)
\label{L5}
\ea
in addition to a tadpole contribution $\L_1=-\phi$ which satisfies the accidental symmetry. Since their discovery, the Galileon have been shown to belong to the more general class of interactions first introduced by Fairlie {\it et. al.}, \cite{Fairlie:1991qe,Fairlie:2011md},  and can be shown to satisfy the recursive relation,
\ba
\label{recursive}
\mathcal{L}_{n+1}\sim-(\p \phi)^2\frac{\delta \mathcal{L}_n}{\delta \phi} \hspace{20pt}{\rm for}\hspace{10pt} n\ge 1\,.
\ea
It is straightforward to check that there cannot be any further invariant beyond $n=5$ in four dimensions because $\delta_\phi \mathcal{L}_5$ is a total derivative.

The phenomenology of the Galileon has been explored in great depth in the literature, and we will review some of their cosmological implications in what follows, but their rich phenomenology relies on the essential fact that they also exhibit a Vainshtein mechanism and can hence be screened at short distances or high energy.  The exact behavior of the field within the strong coupling radius depends on the existence of these higher interactions terms, but the suppression of the force is present for a wide region of parameter space.

\subsubsection{Non-Renormalization Theorem}

As mentioned previously, the presence of interactions at an energy scale $\Lambda \ll \mpl$ is essential for the viability of Galileon theories and for the Vainshtein mechanism to work. Nevertheless these interactions are typically ``irrelevant" from a traditional effective field theory sense and the theory is hence not renormalizable. This comes as no surprises, since gravity itself is not renormalizable, however we are here required to work within a regime where these interactions are important (and even dominate within the strong coupling region when $\p^2 \phi \ll \Lambda^3$). With this in mind, one may expect that the effective field description goes out of control within the strong coupling region and is no longer a valid physical description.

The existence of the Galileon and shift symmetry goes a long way in preventing generic local operators from being generated by loop corrections. However we do still expect all local operators which are invariant under shift and Galileon transformations to be generated at the quantum level.  In particular we expect in principle
\begin{enumerate}
 \item that the Galileon interactions will be themselves renormalized. If correct, this would imply that  the strong coupling scale $\Lambda$ could receive large quantum corrections $\delta \Lambda_{\rm QC}\sim M_\star$, (where $M_\star$ is the cutoff scale, $M_\star\gg \Lambda$) which would hence destroy the entire Galileon framework,
 \item that quantum corrections will also generate operators of the form $\(\p^n \phi\)^m$ and any combinations of these, with $n,m\ge 2$ (which do also respect the Galileon symmetry in a trivial way).
\end{enumerate}

The essence of the non-renormalization theorem formulated in Refs.~\cite{Luty:2003vm} and \cite{Nicolis:2004qq} comes in two parts: 1. The Galileon interactions themselves are not renormalized at all (and hence the scale $\Lambda$ receives no quantum corrections, or in other words the Galileon interactions could be important and radiatively stable) and 2. Higher derivative operators of the form $\(\p^n \phi\)^m$ are generated by quantum corrections, but there exists a regime of interest for the theory, for which they are irrelevant, \ie within the strong coupling region, the field itself can take large values, $\phi \sim \Lambda$, $\p \phi \sim \Lambda^2$, $\p^2 \phi \sim \Lambda^3$, but any further derivative of the field is suppressed, $\p^{n} \phi \ll \Lambda^{n+1}$ for any $n\ge 3$\footnote{This is similar to the situation in DBI scalar field models, where the field operator itself and its velocity is considered to be large $\phi \sim \Lambda$ and $\p \phi \sim \Lambda^2$, but any higher derivatives are suppressed $\p^n \phi \ll \Lambda^{n+1}$ for $n\ge 2$, as shown in section \ref{sec:DBIGalileon}}. In other words, the Effective Field  expansion should be reorganized so that operators which do not give equations of motion with more than two derivatives (\ie Galileon interactions) are considered to be large and ought to be treated  as the relevant operators, while all other interactions (which lead to terms in the equations of motion with more than two derivatives) are treated as irrelevant corrections in the effective field theory language.

To understand in more depth the first point related to the non-renormalization of the Galileon interactions, let us follow the procedure established in \cite{Nicolis:2004qq}. Having the Vainshtein mechanism in mind where the field acquires a large background value, we consider perturbations around an arbitrary background configuration $\phi=\phi_{\rm cl}+\varphi$. Around that background configuration, the second order Lagrangian
for the perturbations is of the form
\ba
\mathcal{L}_{\varphi}=-\frac 12 Z\mn(\phi_{\rm cl})\p^\mu \varphi \p^\nu \varphi\,,
\ea
where the kinetic matrix is symbolically of the form,
\ba
Z \sim 1+c_3 \frac{\p^2 \phi_{\rm cl}}{\Lambda^3}+c_4 \frac{(\p^2 \phi_{\rm cl})^2}{\Lambda^6}+c_5 \frac{(\p^2 \phi_{\rm cl})^3}{\Lambda^9} \,,
\ea
where $c_{3,4,5}$ are the coefficients for the $3^{\rm rd}, 4^{\rm th}$ and $5^{\rm th}$ Galileon interactions.
In order to extract the real physical scale at which interactions arise, it is wise to work in terms of the canonically normalized field $\hat \varphi =\sqrt{Z}\varphi$ which acquires the effective mass
\ba
m_{\rm eff}^2(\phi_c) \sim \frac{(\p Z)^2}{2Z^2}-\frac{\p^2 Z}{Z}\,.
\ea
The Coleman-Weinberg 1-loop effective action is then of the form
\ba
\mathcal{L}_{\rm 1-loop}=m_{\rm eff}^4(\phi_c)\log\frac{m_{\rm eff}^2}{M_\star^2}
\label{CT}
\sim \frac{(\p Z)^4}{Z^4} \log\frac{m_{\rm eff}^2}{M_\star^2}\,,
\ea
where $M_\star$ is the cutoff scale, $M_\star \gg \Lambda$. The second part of the non-renormalization theorem is now more transparent:  even though the field may be large, $\phi_c \sim \Lambda$, $\p^2 \phi_c \sim \Lambda^3$ and hence $Z$ is large, the one-loop effective action only depends on the ratio  $\p Z / Z$. So once again, quantum corrections are under control as long as $\p Z \ll Z$, or in other words as long as $\p^3 \phi_c \sim \Lambda^4$ (and similarly for any higher derivative operator).  Furthermore, if we pay closer attention to the structure of the counterterms arising in the 1-loop effective action, we notice that they all come with at leat one extra derivative as compared to the original interactions, $\p Z \sim \p^3 \phi$.  Thus no counterterms take the Galileon form, and the Galileon interactions are hence not renormalized (the Galileon couplings constants may be tuned to any value and remain radiatively stable, such a tuning is hence technically natural).

Finally, a last ingredient essential for the quantum stability of Galileon theories is the fact that for fluctuations $\hat \varphi$ on top of the background configuration, interactions do not arise at the scale $\Lambda$ but rather at the rescaled strong coupling scale $\hat \Lambda= \sqrt{Z}\Lambda$ which is much larger than $\Lambda$ within the strong coupling region. The higher interactions for fluctuations on top of the background configuration are hence much smaller than expected and their quantum corrections are therefore suppressed. This is the essence of the Vainshtein mechanism itself as presented in section \ref{sec:Vainshtein}, explaining why the interactions with the fluctuations $\hat \varphi$ on top of a background configuration are suppressed in the vicinity of a dense region of space.

\subsection{Massive Gravity}

As seen previously, the decoupling limit of DGP corresponds to a specific realization of the Galileon, which exhibits the Vainshtein mechanism without the presence of ghost. It was subsequently established that this does not only apply to soft massive gravity such as DGP or its higher dimensional extension, but also to hard mass gravity, such as New Massive Gravity in three dimensions, and as we shall see to Ghost-free massive gravity in four dimensions. In what follows we show explicitly how generalizing the Fierz-Pauli theory of massive gravity leads to a Galileon theory in its decoupling limit, and review the procedure underlined in \cite{deRham:2010ik,deRham:2010kj}.

We start by implementing GR with a new potential of the form
\ba
\label{action}
S=\int \d^4 x \sqrt{-g}\, \frac{\mpl^2}{2} \, \(R - m^2\,  \mathcal{U}(H\mn, g\mn)\)\,,
\ea
where $H\mn$ is constructed out of the metric and four \stu scalar fields $\phi^a$,
\ba
\label{Hmn}
H\mn=g\mn-\eta_{ab}\p_\mu \phi^a \p_\nu \phi^b\,,
\ea
such that $H\mn$ represents the fluctuations around flat space-time when taking the unitary gauge $\phi^a=x^a$, $H\mn|_{\phi^a=x^a}=g\mn-\eta\mn\equiv h\mn/\mpl $. As a well-known example, the Fierz-Pauli mass term corresponds to the specific choice of potential, such that in unitary gauge
\ba
\mpl^2 \sqrt{-g}\ \mathcal{U}_{FP}|_{\phi^a=x^a}=-\frac 12 \eta^{\mu\nu}\eta^{\alpha \beta}(h_{\mu \alpha}h_{\nu \beta}-h\mn h_{\alpha \beta})\,.
\ea
The Fierz-Pauli mass term is free of any Boulware-Deser ghost at quadratic order, however it reappears when working at higher order in perturbations (or equivalently when working around a curved background), \cite{Boulware:1973my}. The existence of this Boulware-Deser ghost is most easily seen in terms of the \stu fields, \cite{Deffayet:2005ys}. We start by splitting them into the canonically normalized helicity-1 and -0 modes,
\ba
\phi^a=x^a+\frac{A^a}{\mpl m}+\eta^{ab}\frac{\p_b \phi}{\mpl m^2}\,.
\ea
In what follows we focus solely on the helicity-2 and -0 modes and set $A^a=0$ (this mode does not get sourced at leading order, and setting it to zero is consistent to all orders).
The tensor $H\mn$ is then expressed as
\ba
H\mn=\frac{1}{\mpl} h\mn+\frac{2}{\Lambda^3} \p_\mu \p_\nu \phi-\frac{1}{\Lambda^6}\eta^{\alpha \beta}\, \p_\mu \p_\alpha \phi \p_\nu \p_\beta \phi\,,
\ea
where similarly as in DGP, the strong coupling scale $\Lambda$ is given by $\Lambda=(\mpl m^2)^{1/3}$. It is now clear that for a generic potential term $\mathcal{U}(H\mn)\sim H\mn^n$ includes higher derivatives in $\phi$ of the form  $(\p^2 \phi)^n$ which enter with more than two derivatives at the level of the equations of motion. This implies that more than two initial conditions ought to be provided. This extra information can be interpreted as a physical excitation with the wrong-sign kinetic energy, which in turns implies that the associated Hamiltonian of this theory is unbounded from below (at the quantum level this signals a ghost while at the classical level this corresponds to the Ostrogradski instability), \cite{Ostrogradski,Whittaker,astro-ph/0601672}. This is nothing else but the appearance of the Boulware-Deser (BD) ghost at the non-linear level, \cite{Boulware:1973my,Deffayet:2005ys}.

\subsubsection{Decoupling limit}

To avoid this BD ghost, the potential $\mathcal{U}$ ought to be constructed with great care. When the BD is present, it can already be observed within the decoupling limit, so for simplicity, we will start by constructing a theory which is free of the BD ghost within the decoupling limit, and later show how the absence of ghost holds in all generality. Similarly as in DGP we take the decoupling limit by sending $\mpl \to \infty$ and $m\to 0$ while keeping the scale $\Lambda$ fixed.
It will be useful to construct the following quantity, \cite{deRham:2010kj}
\ba
\K\mupn=\K\mupn(g,H)=\delta\mupn -\sqrt{\delta\mupn - g^{\mu\alpha}H_{\alpha\nu}}\,.
\ea
The tensor $\K$ is built precisely so as to reduce to $\p_\mu\p_\nu \phi$ in the decoupling limit,
\ba
\K\mupn|_{\mpl\to \infty}=\K\mupn|_{h\mn=0}
\equiv \frac{1}{\Lambda^3}\eta^{\mu\alpha}\p_\alpha \p_\nu \phi\,.
\ea
Since self-interactions for $\phi$ arise at the scale $\Lambda$ which remains fixed, while the ones for the helicity-2 enter at the Planck scale, it is straightforward to see that the leading contributions in the decoupling limit can be derived from a simple Taylor expansion,
\ba
m^2 \mpl^2 \sqrt{-g }\, \mathcal{U}(H,g)=\Lambda^3\(\mpl \mathcal{U}|_{h\mn=0}+ h\mn X^{\mu\nu} +\frac{1}{\mpl}h^2\cdots\)\,,
\ea
with
\ba
X^{\mu\nu}=\mpl \frac{\delta }{\delta h\mn} \sqrt{-g} \, \mathcal{U}(H,g)|_{h\mn=0}\,.
\ea
To ensure that the leading contribution is ghost free, we consider the most general following potential
\ba
\label{pot}
\mathcal{U}(g,H)&=&-\([\K]^2-[\K^2]\)-\alpha_3 \([\K]^3-3[\K][\K^2]+2 [\K^3]\) \\
&&-\alpha_4\([\K]^4-6 [\K]^2[\K^2]+3[\K^2]^2+8[\K][\K^3]-6[\K^4]\)\nn\,,
\ea
where square brackets represent the trace of a tensor, $[T]\equiv  T^\mu _{\, \mu}$, $[T^2]\equiv T\mupn T^\nu_{\, \mu}$, etc\ldots \,.
This potential is specifically built such that its leading contribution in the decoupling limit $\mathcal{U}|_{h\mn=0}$ reduces to a total derivative\footnote{One could also consider an additional linear contribution of the form $[\K]$ but this is actually a redundant operator, the tadpole can always be reabsorbed by field redefinition of the dynamical metric.}. To see the first interactions arising in the decoupling limit we then need to work at first order in $h\mn$. As derived in \cite{deRham:2010ik,deRham:2010kj}, the tensor $X\mn$ associated to the potential \eqref{pot} is expressed by
\ba
X\mn=\frac{1}{\Lambda^3} X\mn^{(1)}-\frac{1+3\alpha_3}{\Lambda^6} X\mn^{(2)}+\frac{3(\alpha_3+4\alpha_4)}{\Lambda^9} X\mn^{(3)}\,,
\ea
with
\ba
X\mn^{(2)}&=&\Phi\mn-[\Phi] \eta\mn\\
X\mn^{(3)}&=&\Phi^2\mn-[\Phi] \Phi\mn-\frac 12 ([\Phi]^2-[\Phi^2])\eta\mn \\
X\mn^{(4)}&=&\Phi^3\mn -[\Phi]\Phi^2\mn +\frac 12 ([\Phi]^2-[\Phi^2])\Phi\mn
-\frac 16 ([\Phi]^3-3[\Phi^2][\Phi]+2 [\Phi^3])\eta\mn
\,,
\ea
with $\Phi\mn=\p_\mu\p_\nu \phi$. Although this seems to differ from a Galileon model at first sight, all the Galileon properties enumerated in section \ref{DGP} are equivalently satisfied here. This comes actually as no surprise as the resulting action for the field $\phi$ is expressible in terms of Galileon interactions after appropriate redefinition of $h\mn$. Upon the field redefinition
\ba
h\mn = \bar h\mn+2\phi \eta\mn -\frac{1+3\alpha_3}{2\Lambda^3} \p_\mu \phi \p_\nu \phi\,,
\ea
the resulting decoupling Lagrangian for gravity amended with the potential \eqref{pot} is given by
\ba
\L&=&-\frac 14 \hat{h}^{\alpha\beta}\hat{\mathcal{E}}^{\mu\nu}_{\alpha \beta}\hat{h}\mn
+3  \phi \Box \phi -\frac{3(1+3\alpha_3)}{2\Lambda^3}(\p \phi)^2 \Box \phi\\
&+&\frac{1}{\Lambda^6}\(\frac 18 (1+3\alpha_3)^2-\frac 12 (\alpha_3+4\alpha_4)\)(\p \phi)^2\([\Phi^2]-[\Phi]^2\)
-\frac{\alpha_3+4\alpha_4}{2\Lambda^6}\hat h^{\mu\nu}X^{(3)}\mn \nn\\
&-&\frac{5}{16\Lambda^9}(1+3 \alpha_3)(\alpha_3+4 \alpha_4)(\p \phi)^2
\([\Phi]^3-3[\Phi][\Phi]^2+2[\Phi^3]\)\,,\nn
\ea
where $\hat{\mathcal{E}}$ is the linearized Einstein tensor,
$\hat{\mathcal{E}}^{\alpha\beta}\mn h_{\alpha\beta}=\Box h\mn-\p_\alpha \p_{(\mu}h^\alpha_{\nu)}+\p_\mu\p_\nu h-\eta\mn (\Box h-\p_\alpha\p_\beta h^{\alpha\beta})$. We see in particular that when $\alpha_3+4\alpha_4=0$ we recover precisely the Galileon set of interactions (with now a coupling to matter which would be of the form $\phi T +\frac{1+\alpha_3}{4\Lambda^3}\p_\mu \phi \p_\nu \phi T^{\mu\nu}$), and the Vainshtein mechanism can be seen to work for a large range of parameters, even beyond the decoupling limit, \cite{Koyama:2011yg,Chkareuli:2011te}, see also ref.~\cite{Iorio:2012pv}.

In the case where $\alpha_3+4\alpha_4\ne 0$, the helicity-2 and helicity-0 modes are not generically diagonalizable in a covariant way. As we shall see below, this has interesting consequences for cosmology and represents a new coupling which is not encoded by the Galileon family of interactions.

\subsubsection{Beyond the decoupling limit}
The previous subsection focused on how to construct a theory of massive gravity which is ghost-free in the decoupling limit, which may only be used to describe physical processes that happen at most at the energy scale $\Lambda$. Since this scale is very small, (typically $\Lambda \sim 10^{-40}\mpl$), one really ought to understand the consistency of the theory beyond this decoupling limit. Fortunately we have derived the potential for the graviton in a completely covariant way, and so the action \eqref{action} for massive gravity, implemented with the potential \eqref{pot}, where $H\mn$ is given in terms of the four \stu fields as in \eqref{Hmn} remains a fully valid description beyond the decoupling limit (the only difference when working beyond the decoupling limit is that higher order interactions with $h\mn$ may no longer be omitted).

To explore the phenomenology of this theory beyond the decoupling limit (\ie at energy scales above $\Lambda$), one can of course keep the same language as what was used within the decoupling limit and split the degrees of freedom into two modes present in the metric $g\mn=\eta\mn+h\mn$, and the other modes included in the \stu fields $\phi^a = x^a +V^a$.
Even though valid, this language turns out not to give the best description of the system. The reason for this is that the quantity $\phi$ that appears in the \stu field as $V^a=\p^a \phi$ does no longer fully describe the helicity-0 mode of the graviton when interactions with gravity are included, \cite{deRham:2011rn,deRham:2011qq}.

To be more precise, when ignoring gravity (\ie when working around flat space-time, which is effectively what happens within the decoupling limit since all higher interactions in $h$ are suppressed in the limit $\mpl \to \infty$), space-time enjoys a global Lorentz invariance $x^a \to \Lambda^a_b x^b$. Moreover, the theory is also invariance under an independent internal transformation of the four \stu fields, which is also a global Lorentz transformation $\phi^a \to \tilde \Lambda^a_b \phi^b$. Thus within the decoupling limit, one may work in the representation of the single group $\tilde \Lambda = \Lambda$, and the \stu fields hence behave as vectors under this global Lorentz symmetry. It does therefore makes sense to perform a scalar-vector-tensor decomposition of the fields in this representation, and express $\phi^a=x^a+A^a+\p^a \phi$ in that limit, where $\phi$ then captures the physics of the helicity-0 mode. However beyond the decoupling limit space-time loses the global Lorentz symmetry and $\phi$ no longer captures the relevant degree of freedom.

When maintaining the split $\phi^a=x^a+A^a+\p^a \phi$ beyond the decoupling limit, then $\phi$ will maintain its Galileon symmetry in a manifest way (simply by construction), but this will not be the case for the helicity-0 mode itself. Furthermore beyond the decoupling limit, $\phi$ may appear with higher derivatives in its equations of motion, however once again, since $\phi$ bears no explicit physical meaning beyond the decoupling limit (no physical processes involve only scatterings of $\phi$ with itself),
this does not imply the presence of ghost. Indeed, one can show explicitly that unitarity is still preserved,  \cite{deRham:2011rn,deRham:2011qq,Hassan:2012qv}. However to see this more explicitly, it is wiser to work in the Hamiltonian language, and to work in unitary gauge $\phi^a=x^a$ so that all the degrees of freedom are then carried by the metric. This analysis was  successfully performed to fourth in perturbations in the Hamiltonian in Ref.~\cite{deRham:2010kj}, and then generalized to all orders in \cite{Hassan:2011hr}. The existence of a secondary constraint was then demonstrated in \cite{Hassan:2011ea}. Recent work in Ref.~\cite{Mirbabayi:2011aa} also clarifies how the absence of ghost in the decoupling limit generalizes beyond that limit.

\subsection{New Massive Gravity}

Whilst a massless spin-2 field propagates no dynamical degree of freedom in three dimensions, the same does not hold for a massive spin-2 field which can excite two polarizations, one of which behaves as a helicity-0 mode. This realization has made the study of three-dimensional massive gravity especially appealing. One explicit realization of massive gravity in three dimensions was proposed a few years ago by in \cite{Bergshoeff:2009hq} and relies on the introduction of higher curvature terms. Such terms usually lead to new dofs with the opposite sign kinetic term and would be unacceptable in most cases, however since massless spin-2 field do not propagate any physical dof, one can ``survive" in three dimensions with a wrong-sign massless spin-2 field without further ado. Whilst the  situation is very specific to three dimensions and cannot be extended to higher dimensions in an straightforward manner, the resulting decoupling limit theory carries very similar features to what is obtained in four-dimensional theories of massive gravity (see Ref.~\cite{Bergshoeff:2012ud} for a recent attempts at the linear level, based on Refs.~\cite{Curtright:1980yk,Curtright:1980yj}). To see this explicitly, we start with the original formulation of NMG,
\ba
S_{\rm NMG}=\int \d^3 x \mpl \sqrt{-g}\(-R+\frac{1}{m^2}(R\mn^2-\frac 38 R^2)\)\,.
\ea
To see the Galileon structure arising, let us focus on the conformal mode, $g\mn  =\(1+ \frac{\phi}{\sqrt{\mpl}} \) \eta\mn $ and consider the decoupling limit $\mpl \to \infty$, whilst keeping the scale $\Lambda_{3d}=(m^2 \sqrt{\mpl})^{2/5}$ fixed. In that limit, the resulting action for NMG is then precisely of a Galileon form, \cite{deRham:2011ca}
\ba
\L_{\rm NMG}^{(\rm dec)}=2\phi \Box \phi -\frac{1}{2\Lambda_{3d}^{5/2}}(\p\phi)^2 \Box \phi\,.
\ea
Thus in all known models of massive gravity be it DGP, New Massive Gravity or the Ghost-free theory of massive gravity in four-dimensions, the absence of the BD ghost, the decoupling limit resembles that of a Galileon theory. Whilst these models represent the natural embedding of Galileon scalar fields beyond the decoupling limit, we shall see in what follows another class of braneworld configuration from where Galileons arise when considering a non-relativistic limit.

\subsection{DBI - Galileon}
\label{sec:DBIGalileon}
Some of the features of the Galileon are very similar in spirit to the DBI world action for a brane embedded within a wrapped extra dimension. In particular, DBI inflation relies on a braneworld action of the form
\ba
\L_{\rm DBI} &=& f(\phi)^4\(1-\sqrt{1+f(\phi)^{-4}(\p \phi)^2}\)-V(\phi)\\
&=& -\frac 12 (\p \phi)^2 + \frac{1}{6 f(\phi)^4}(\p \phi)^4+\cdots -V(\phi)\,,
\ea
where dimension-8 and other higher irrelevant operators such $(\p \phi)^4$ are important, $\p \phi \lesssim f(\phi)^2$ without yet going beyond the regime of validity of the theory. The reason for that is here again due to the existence of a non-renormalization theorem, whereby a class of irrelevant operators may be important at energy scales much lower than the Planck scale $\p \phi \sim f^2 \ll \mpl^2$ but involve quantum corrections which themselves are negligible as long as $\p^2 \phi \ll f^3$. These features is very similar to the Galileon, and as we shall see below, both theories can actually be ``reunited" within a common braneworld framework as shown in Ref.~\cite{deRham:2010eu}.

To see this explicitly, let us consider a four-dimensional brane, embedded within a five-dimensional maximally symmetric extra dimension (we focus on flat extra dimension within this review, although this framework is easily generalizable to more general geometries). The Lovelock invariants have already been established almost half a century ago \cite{Lovelock:1971yv} as providing all the possible geometrical quantities which are free of any ghostly additional degree of freedom. In four dimensions, there are two of them, namely the cosmological constant and the Einstein-Hilbert scalar curvature term, leading to the usual Einstein equation. We can therefore start by considering these two contributions on the four dimensional brane. In five dimensions, in addition to these two geometrical contributions, the Gauss-Bonnet term represent a non-trivial higher-derivative term which combines the curvature terms in such a way so as to remain second order at the level of the equations of motion. If present within the fifth dimension, the Ricci scalar curvature and the Gauss-Bonnet term induce specific boundary terms on the brane similarly as the Gibbon-Hawking boundary term.
 In addition to these four geometrical contributions, we can also include a tadpole term arising from the integration over the bulk loops, similarly as in DBI.
Considering all these geometrical quantities, the most general four-dimensional action arising from all the possible five and four-dimensional Lovelock invariant is thus of the form
\ba
S=\mpl^2 \int \d^4 x \sqrt{-g}\(\L_2+c_3\L_3+c_4\L_4 +c_5\L_5\)\,,
\ea
where $\L_2$ is the brane tension or four-dimensional cosmological constant which we choose to write in terms of a scale $m$ of dimension mass and a contribution from the tadpole,and $\L_{3,4,5}$ are respectively the contributions for the extrinsic curvature, the four-dimensional Einstein-Hilbert term and the boundary term associated with Gauss-Bonnet  curvature in the bulk,
\ba
\L_2&=&m^2 \(\frac{1}{\sqrt{-g}}-1\)\\
\L_3&=&  m \,  K\\
\L_4&=&    R\\
\L_5&=& m^{-1}\,  \K_{\rm GB}\,.
\ea
We consider a flat bulk with a four-dimensional brane localised at $y=\hat \phi \equiv \phi/\mpl m$.
The induced metric on the brane is then given  $g\mn=\eta\mn+ \p_\mu \hat \phi \p_\nu \hat \phi$, leading to the extrinsic curvature on the brane,
\ba
K\mn =- \frac{ \p_\mu \p_\nu \hat \phi}{\sqrt{1+(\p \hat \phi)^2}} \,,
\ea
whilst the boundary term associated with the Gauss-Bonnet curvature in the bulk is \cite{Davis:2002gn}:
\ba
\mathcal{K}_{\rm GB}=-\frac 23 K\mn^3+K K\mn^2-\frac 13 K^3-2G\mn K^{\mu\nu}\,,
\ea
where $G\mn$ is the Einstein tensor associated with the induced metric $g\mn$. The exact expression for these different invariants was provided in Ref.~\cite{deRham:2010eu}. Since the brane is embedded within a flat fifth dimension, the four-dimensional action should be invariant under translation and boosts along the fifth dimension, which applied on the brane position, corresponds to the infinitesimal global symmetry,
\ba
\phi \to \phi + c + v_\mu x^\mu + \frac{1}{m^2 \mpl^2}\phi \, v^\mu \p_\mu \phi\,.
\ea
We now consider the non-relativistic limit, $(\p \hat \phi) \to 0$, which can be achieved, by sending once again $\mpl \to \infty$ while keeping the scale $\Lambda=(\mpl m^2)^{1/3}$ fixed. The resulting action on the brane is then nothing else than the four Galileon interactions proposed in (\ref{L2}-\ref{L5}), with the associated non-relativistic global Galileon symmetry:
\ba
\phi \to \phi + c + v_\mu x^\mu\,.
\ea
A analogue construction of the Galileon interactions from five-dimensional Lovelock invariants was also later developed using a compactified extra dimension \`a la Kaluza-Klein, \cite{VanAcoleyen:2011mj}.

In all of these models, be it braneworld models or hard or soft massive gravity, the scalar new degree of freedom which encodes either the helicity-0 mode or the brane-bending mode behaves as a Galileon scalar field within a specific ``decoupling" limit. When expressed in terms of a Galileon scalar field, the Vainshtein mechanism is most easily tractable, and all these models enjoy a healthy strong coupling effect, at least for some choice of their parameters. This strong coupling is essential in understanding how such a light scalar mode could be present and remain as yet unobserved. In what follows we consider different cosmological applications of the Galileon and show how it can play an important role for the late-time acceleration of the Universe as well as for inflation.


\section{Galileon Cosmology}
\label{sec:cosmology}

\subsection{Self-acceleration in DGP and Galileons}

One of the breakthrough of  the DGP model, \cite{Dvali:2000hr}, was the realization that the Universe could be accelerating without the presence of any cosmological constant or other  source of ``dark-energy" but sourced by the graviton very own dofs and in particular by its helicity-0 mode, \cite{Cedric}. Within this ``self-accelerating" solution of DGP, the expansion of the Universe is naturally accelerating without the need of any external source or cosmological constant. It was quickly realized though that this branch of solution contained a negative norm state which was hence propagating a ghost. To see this explicitly, let us consider the decoupling limit of DGP which takes the form (including the helicity-2 mode),
\ba
\L_{\rm DGP}^{(\rm dec)}=-\frac 14 \bar h^{\alpha \beta}\hat{\mathcal{E}}^{\mu\nu}_{\alpha \beta}\bar h\mn+3 \phi \Box \phi\pm\frac{1}{\Lambda^3}(\p \phi)^2 \Box \phi+\frac{1}{2\mpl} h\mn T^{\mu\nu}+\frac{2}{\mpl}\phi T\,,
\ea
where $\bar h\mn$ and $\phi$ represents the helicity-2 and -0 modes respectively, and the linearized induced metric on the brane in DGP is given by $g\mn=\eta\mn+\frac{1}{\mpl}\(\bar h\mn+2\phi \eta\mn\)$. The `$-$'-sign corresponds to the normal (stable) branch of DGP while the `$+$'-branch is the self-accelerating one.  This leads to the standard  Einstein's equations for the helicity-2,
\ba
\mpl \delta G\mn= \hat{\mathcal{E}}^{\alpha\beta}\mn \bar h_{\alpha \beta} =\frac{1}{\mpl}T\mn\,,
\ea
such that in the absence of any source the helicity-2 mode is not excited ($\bar h\mn=0$ up to gravitational waves). For the helicity-0 mode on the other hand, there is a non-trivial configuration even in the absence of external sources,
\ba
\label{DGP_eom}
3\Box \phi\mp\frac 1{\Lambda^3} \((\Box \phi)^2-(\p_\mu \p_\nu \phi)^2\)=-\frac{1}{\mpl}T\,.
\ea
The homogeneous configuration $\p^2 \phi=0$ is always a solution when $T=0$, but there exists also another branch which admits $\phi_0 = \pm \frac 12 \Lambda^3 x_\mu x^\mu$ as a solution. For this branch of solution, the linearized induced metric on the brane is of the form $g\mn = (1\pm \Lambda^3 x^2) \eta\mn$, which is nothing else but (anti) de Sitter linearized around Minkowski. In particular when working with the `$-$'-sign, one can only achieve a Minkowski or an AdS solution, while the `$+$'-sign allows for a de Sitter solution.   This goes to show how the induced metric on a DGP brane can be de Sitter even in the absence of any cosmological constant or dark-energy.   However one can easily notice that fluctuations on the top of this self-accelerating solution bear no kinetic term, already signaling the presence of strong coupling issues, on top of the self-accelerating branch. This issue is resolved when including any external source, be it radiation, dark matter or other typically present in the Universe, but it can then be shown that the helicity-1 mode of the graviton then bears a negative kinetic term hence signaling the existence of a ghost, \cite{Koyama:2005tx}.

The presence of a ghost within the self-accelerating branch of DGP can be avoided when working  in a more general Galileon framework, developed in Ref.~\cite{Nicolis:2008in} (or other extensions of DGP, \cite{deRham:2006pe}). The equation of motion for the helicity-0 mode \eqref{DGP_eom} then gets replaced by
\ba
c_1\Box \phi+\frac{c_2}{\Lambda^3}  \([\Phi]^2-[\Phi^2]\)+\frac{c_3}{\Lambda^6} \([\Phi]^3-3[\Phi^2][\Phi]+2 [\Phi^3]\)&& \nn \\
+\frac{c_4}{\Lambda^9} \([\Phi]^4-6 [\Phi]^2[\Phi^2]+3[\Phi^2]^2+8[\Phi][\Phi^3]-6[\Phi^4]\)
&=&-\frac{1}{\mpl}T\,,
\ea
so there are non-trivial branches of solution  in the absence of matter $T=0$, for different choices of parameters $c_{2,3,4}$. In particular there exists a region in parameter space for which fluctuations around the self-accelerating solution are stable and exhibit the Vainshtein mechanism, \cite{Nicolis:2008in}.

In what follows, we focus instead on a slightly different realization of self-accelerating solutions which relies on the new specific interactions that arise within massive gravity.

\subsection{The accelerating Universe in Massive Gravity}

To study the self-accelerating solutions in massive gravity, we review the approach provided in \cite{deRham:2010tw}.
In this case it is more convenient to work in Jordan frame  where the linearized metric is given by $g\mn = \eta\mn+h\mn / \mpl$ and $h\mn$ is sourced both by matter and the helicity-0 mode,
\ba
\label{Einstein_MG}
\mpl \delta G\mn= \hat{\mathcal{E}}^{\alpha\beta}\mn \bar h_{\alpha \beta} =\frac{1}{\mpl}\(T\mn+T^{(\pi)}\mn\)\,,
\ea
where $T^{(\pi)}\mn$ is the additional contribution arising from the helicity-0 mode,
\ba
T^{(\pi)}\mn= \Lambda^3X\mn\,.
\ea
The equation of motion for $\phi$ is given by:
\ba
\label{eq_phi}
\p_\alpha \p_\beta h\mn \, \varepsilon^{\mu \alpha \rho \sigma}\(-\frac 12 \varepsilon^{\nu\beta}{}_{\rho \sigma}+\frac{1+3 \alpha_3}{2 \Lambda^3}
\varepsilon^{\nu\beta\gamma}{}_{ \sigma} \Phi_{\rho \gamma}+\frac{3(\alpha_3+4 \alpha_4)}{2\Lambda^6}\varepsilon^{\nu\beta\gamma\delta} \Phi_{\rho \gamma}\Phi_{\sigma \delta}\)=0\,,
\ea
where $\varepsilon_{\mu\nu\alpha\beta}$ is the Levi-Cevita tensor.
This has two important consequences, first of all the existence of an additional $T^{(\pi)}\mn$ implies the existence of self-accelerating solutions in the absence of matter, $T\mn=0$, as along as $T^{(\pi)}\mn\sim g\mn$.  However, another peculiarity of this framework is the existence of degravitating or screening solutions, where the helicity-0 mode can absorb the contribution from a cosmological constant, $T\mn=-\lambda_{\rm CC} g\mn$ and $T^{(\pi)}\mn=\lambda_{\rm CC} g\mn$, in such a way that the geometry remains flat ($g\mn=\eta\mn$) even in the presence of a potentially large cosmological constant. Such a solution would not be possible within a pure Galileon framework and the difference originates from the additional coupling to matter $\p_\mu \phi \p_\nu \phi T^{\mu\nu}$ that arises when switching to the Einstein frame. As we shall see below such terms can also have important observational signatures.

\subsubsection{Self-accelerating Branch}
We start by giving a brief overview of the self-accelerating solution, before moving onto the degravitating branch. We focus on linearized solutions around flat space-time and look for configurations of the form $h\mn = -\frac 12 H^2 x^2 \eta\mn$, where $H$ is the equivalent of the Hubble parameter and $\phi=\frac 12 \Lambda^3 q x^2$. This branch corresponds to the case where the constant $q$ is related to the Hubble parameter by \eqref{eq_phi}
\ba
\label{q_1}
H^2\(-\frac{1}{2}+2a_2 q+3 a_3 q^2\)=0\,,
\ea
with $a_2=-\frac 12 (1+3\alpha_3)$ and $a_3=\frac 12 (\alpha_3+4\alpha_4)$,
and the Einstein equation in the absence of external sources is given by \eqref{Einstein_MG}
\ba
\label{q_2}
\mpl H^2=2 q \Lambda^3  \( -\frac12+   a_2 q+a_3 q^2 \)\,.
\ea
Self-accelerating solutions for which $H>0$ thus exist for a large family of parameters $a_{2,3}$. To investigate  this branch further, we analyze small perturbations for the helicity-2 and -0 modes, $h\mn=-\frac 12 H^2 x^2 \eta\mn+\chi\mn$ and $\phi=\frac 12 \Lambda^3 q x^2+\varphi$. The resulting Lagrangian for the perturbations (up to a total derivative) reads as follows \cite{deRham:2010tw}
\ba
\L&=& -\frac{1}{2}\chi^{\mu\nu}\hat{\mathcal{E}}^{\alpha\beta}_{\mu\nu}
\chi_{\alpha\beta}+6(a_2+3a_3q)\frac{H^2 \mpl}{\Lambda^3}\varphi
\Box\varphi-3a_3\frac{H^2 \mpl}{\Lambda^6}(\partial\varphi)^2\Box\varphi \nn \\
&+&\frac{a_2+3 a_3 q}{\Lambda^3}\chi^{\mu\nu} X^{(2)}_{\mu\nu}[\varphi]
+\frac{a_3}{\Lambda^6}\chi^{\mu\nu} X^{(3)}_{\mu\nu}[\varphi]+\frac{\chi^{\mu\nu}T\mn}{\mpl}\,.
\label{PertL}
\ea
As mentioned in \cite{deRham:2010tw}, a key feature of this self-accelerating branch is the fact that there is no quadratic mixing between
$\chi$ and $\varphi$, which implies that $\chi$ identifies the correctly diagonalized helicity-2 mode at quadratic order and only the helicity-2 mode couples to external sources $T\mn$ at that level. Thus on top of the self-accelerating solution, the helicity-0 mode does not couple to external sources at quadratic order and will only do so at higher order (through a coupling of the form $\p_\mu \varphi \p_\nu \varphi T^{\mu\nu}$). This results remains unaffected by the presence of cosmological matter at the background level.
Therefore, for arbitrary external sources, it is always consistent within the decoupling limit to keep the fluctuation of the helicity-0 unexcited, $\varphi=0$.
This mechanism is a unique feature of the self-accelerating branch in Ghost-free massive gravity. Beyond the decoupling limit we expect the coupling between the helicity-0 and -2 mode to re-emerge (and hence a coupling of the helicity-0 mode to matter), however such a coupling ought to be suppressed by additional powers of $m$, and should therefore only be relevant a very large distance scales.

From the perturbed Lagrangian \eqref{PertL}, we see directly that the situation is quite different from DGP. In particular even though the helicity-2 and -0 modes do not couple at the linearized level, the helicity-0 mode keeps a non-trivial kinetic term proportional to $a_2+3a_3q$. Fluctuations around the self-accelerating branch are hence healthy (no ghost nor tachyon) if $a_2+3a_3 q>0$. This condition, combined with the background equations of motion \eqref{q_1} and \eqref{q_2} and the requirement that $H^2>0$ are satisfied simultaneously if the parameters $a_{2,3}$ satisfy \cite{deRham:2010tw}
\ba
a_2<0 \hspace{10pt}{\rm and}\hspace{10pt}-\frac{2a_2^2}{3}<a_3<-\frac{a_2^2}{2}\,.
\label{bounds}
\ea
In particular we see that the presence of the third tensor $X^{(3)}\mn$ (which has direct equivalent in pure Galileon theories) is essential for the stability of this self-accelerating branch. The resulting Hubble parameter is then of the order of the graviton mass,
\ba
H^2= m^2[2 a_2 q^2+2 a_3q^3-q]>0,\hspace{10pt}{\rm with}\hspace{10pt}q=-\frac{a_2}{3a_3}+
\frac{(2a_2^2+3a_3)^{1/2}}{3\sqrt{2}a_3}\,.
\ea
Further investigations of this self-accelerating branch have established that the helicity-1 mode is likely to become unhealthy around this configuration unless it remains unexcited, \cite{Koyama:2011wx}. This would have serious implications for the self-accelerating branch which should be understood further. We now turn to the other branch of solution which allows the degravitation of the cosmological constant.

\subsubsection{Degravitating branch}
An orthogonal approach to self-acceleration, is to address the question of whether a large cosmological constant (vacuum energy) could be present, whilst only given rise to a very weak late-time acceleration of the Universe. This is the idea behind degravitation, \cite{degravitation}. A first step in the study of degravitation is to understand whether there can exist a static solution (Minkowski space-time) even in the presence of a cosmological constant, which could be a late-time attractor for the Universe. One way to achieve this within massive gravity is to ``absorb" the effect of the cosmological constant through the helicity-0 mode. We review here again the framework derived in \cite{deRham:2010tw}, and consider a cosmological constant,  $T\mn=-\lambda_{\rm CC} g\mn$ absorbed by the helicity-0 mode $T^{(\pi)}\mn=\lambda_{\rm CC} g\mn$ in such a way that the resulting curvature vanishes, $G\mn=0$. Using the same notation as for the self-accelerating branch, we hence set $H=0$ for which \eqref{q_1} is then trivially satisfied and the helicity-0 mode satisfies
\ba
\label{pi0Jordan}
-\frac 12 q+ a_2 q^2+a_3 q^3=-\frac{\lambda_{\rm CC}}{6 \Lambda^3 \mpl}\,.
\ea
As long as the parameter $a_3$ is present ($\alpha_3+4\alpha_4\ne 0$), Eq.~\eqref{pi0Jordan} has always
at least one real root. There is therefore a flat solution for an arbitrarily large
cosmological constant. A further analysis of the scalar, vector and tensor fluctuations shows that this branch of solution if it exists is always stable.

However a drawback of this degravitating solution is that the resulting scale for helicity-0 interactions are no longer governed by the parameter
$\Lambda$, but rather by the scale determined by the cosmological constant to be absorbed,
$\tilde \Lambda\sim (\lambda_{\rm CC}/\mpl)^{1/3}$. To see this, let us pursue the analysis
of the fluctuations around the degravitating branch and keep
the higher order interactions. The resulting Lagrangian involving the helicity-0 mode is then
\ba
\mathcal{L}^{(2)}=-\frac 12  h^{\mu\nu}\(X^{(1)}\mn[\varphi]+\frac{\tilde a_2}{\tilde \Lambda^3}X^{(2)}\mn[\varphi]
+\frac{\tilde a_3}{\tilde \Lambda^6}X^{(3)}\mn[\varphi]\)\,,
\ea
with
\ba
\frac{\tilde a_2}{\tilde \Lambda^3}&=&-2\frac{a_2+3a_3 q}{\Lambda^3 (-1+4 a_2 q+6 a_3 q^2)^2}
\sim \frac{\mpl}{\lambda_{\rm CC}}\\
\frac{\tilde a_3}{\tilde \Lambda^6}&=&-\frac{2a_3}
{\Lambda^6 (-1+4 a_2 q+6 a_3 q^2)^3}\sim \(\frac{\mpl}{\lambda_{\rm CC}}\)^2\,,
\ea
assuming $a_{2,3}\sim \mathcal{O}(1)$. Fifth force constraints usually requires interactions to kick in at least at a scale
${ \tilde \Lambda^3/\mpl} \lesssim (10^{-33}~{\rm eV})^2$ (this bound might be relaxed slightly in the presence of $X^{(3)}$),  which implies that the
the maximum allowed value of vacuum energy  that
can be screened without being in conflict with observations is fairly low,
of the order of  $(10^{-3}~{\rm eV})^4$, (unless the coefficients $a_{2,3}$ are tuned to different natural values, which would still remain technically natural). Although not phenomenologically viable, the existence of this degravitating solution opens the door for a new set of possibilities to tackle the cosmological constant problem. This degravitating solution can also see an analogue within the full theory beyond the decoupling limit, \cite{D'Amico:2011jj}. See Refs.~\cite{Chamseddine:2011bu} for further work in massive gravity with an open FRW cosmology, Ref.~\cite{Koyama:2011xz} for self-accelerating solutions in the full massive gravity theory,
Refs.~\cite{Hassan:2011zd} for cosmology in bi-gravity theories and Refs.~\cite{deRham:2011by} for cosmology in massive gravity beyond the decoupling limit.

\subsection{Galileon Inflation}

As seen in the previous section, the Galileon can play an essential role for cosmology at a time where the Hubble parameter is of the order of $H\sim (\Lambda^3/\mpl)^{1/2}$, such that for very low strong coupling scale $\Lambda\sim 10^{-13}$eV the Galileon affects the late evolution of the Universe, whereas if the Galileon interactions arise at a much larger strong scale, they could have an important impact during inflation (in that case the Galileon should then decay during reheating explaining why it is not observed today, since a larger strong coupling scale would not allow for a sufficiently strong Vainshtein mechanism at late time). This philosophy has received a great attention in the past few years, and one cannot within this short review do justice to all the advances made in that direction. We therefore only focus on the model of Galileon Inflation presented in \cite{Burrage:2010cu}, and refer to \cite{Kobayashi:2010cm} for further work on Galileon Inflation, G-Inflation and the generation of non-gaussianities in these classes of models and to Refs.~\cite{Creminelli:2010ba,LevasseurPerreault:2011mw} for work on Galilean Genesis.

The idea behind Galileon Inflation is to rely on the non-renormalization theorem of the Galileon class of interactions to  build a radiatively stable model of inflation. We hence consider the Galileon to play the role of the usual inflaton scalar field, with  all the Galileon interactions (or their covariant counterpart around an FRW background, \cite{Deffayet:2009wt,Deffayet:2009mn,deRham:2010eu}) arising at the scale $H\ll \Lambda\ll \mpl$ (where $H$ is the Hubble parameter during inflation) in addition to the tadpole contribution $\lambda \phi$, which satisfies the same properties as the other Galileon interactions. All these operators satisfy the shift symmetry which is essential for the production of a scale invariant power spectrum, but as such would not allow for a graceful exit of inflation. However in addition to these operators one may also consider a pure mass term $m^2 \phi^2$ which breaks the shift symmetry but does affect the non-renormalization theorem in any way, \cite{Burrage:2010cu}. The stability of the effective field theory under radiative corrections is thus preserved in the presence of this mass term, which now allows for a graceful exit of inflation.

These considerations are technically correct only in flat space-time while the coupling with gravity breaks the Galilean symmetry, which in turn leads to additional corrections. However these corrections are suppressed by at least two powers of $(H/\Lambda)\ll 1$ and can thus be neglected during inflation.

In the case where the Galileon interactions are negligible $Z=H \dot \phi /\Lambda^3 \ll 1$, the perturbations are weakly coupled and one recovers a standard slow-roll inflation scenario. However Galileon Inflation also allows for another regime $Z\gtrsim 1$, where the interactions can be important without yet going beyond the regime of validity of the Effective Field Theory. This is the regime we are interested in, and the cosmological background evolution is then determined by the largest Galileon operator (assuming all the Galileon interactions arise at the same coupling scale $\Lambda$).

Focusing the study on the effect of fluctuations around this cosmological background
we follow small fluctuations $\xi$, constructed from the background solution by mapping $t \mapsto t + \xi(\vec{x},t)$ on a hypersurface of constant time and  work in a gauge where such hypersurfaces are spatially flat. Up to cubic order in these fluctuations, the resulting Lagrangian then includes terms of the form,
\ba
\L_{\xi} \supseteq a(t)^3 \Bigg[
			\alpha(t) \left\{ \dot{\xi}^2 - \frac{c_s(t)^2}{a(t)^2} (\partial \xi)^2
			\right\}
			+ g_1(t) \dot{\xi}^3
			+ \frac{g_2(t)}{a(t)^2} \dot{\xi} ( \partial \xi )^2
			+ \frac{g_3(t)}{a(t)^4} ( \partial \xi )^2 \partial^2 \xi
		\Bigg] ,
	\label{eq:phi-action}
\ea
where $a(t)$ is the scale factor and $g_{1,2,3}$ as well as $\alpha(t)$ and $c_s(t)$ depend on the background field configuration and can vary over a relatively large range of scales depending on the exact value of the dimensionless scale $Z$.
The operators with coupling $g_1$ and $g_2$ are dimension-6 operators, while the last one with coupling $g_3$ is a dimension-7 operator. In a standard field theory, as well as for DBI inflation, one cannot rely on the dimension-7 operator to be important without going beyond the regime of validity of the Effective Field Theory. In the case of Galileon Inflation, the situation is however different, and the operator-7 can well be of the same order as the others if the dimension-6 operators are suppressed by additional powers of $H/\Lambda$. In the case where the dimension-7 operator is non-negligible, the structure of the non-gaussianities is much richer than what is present in other standard models of slow-roll inflation or even DBI inflation. In particular while the two dimension-6 operators $\dot{\xi}^3$ and $\dot{\xi} ( \partial \xi )^2$ typically lead to a contribution to the non-gaussianity parameter $f^{(g_1)}_{NL}\sim g_1/\alpha$ and similarly the operator $f^{(g_2)}_{NL}\sim g_2/\alpha c_s^2$ respectively, the dimension-7 operator is suppressed by even more powers of the sound speed, $f^{(g_3)}_{NL}\sim g_3/\alpha c_s^4$. DBI inflation relies on the second contribution which is already suppressed by two powers of the sound speed. In the case of Galileon Inflation, the amount of non-gaussianity could be of the same order or even higher, but carried by a different set of operators and hence leading to different shapes of triangle.

Whilst the exact predictions of this model are quite sensitive to the precise value of the different parameters, it opens up the possibility of a new class of non-gaussianities. This possibility was then further investigated by \cite{Creminelli:2010qf} if the leading interactions from operators of the form $(\p \phi)^2 (\p^2 \phi)^n$ vanish, the next to leading order operators of the form $(\p^2 \phi)^n$ become the dominant ones. Dimension-9 operators can then lead to important contributions to the non-gaussianities and arise with their own specific shape. More precisely, two of these operators can lead to equilateral triangles whilst another operator leads to a flattened isocele triangle.

\section{Outlook}

Infrared modified theories of gravity typically come with new degrees of freedom which manifest themselves at low-energy. At high energy, on the other hand we expect to recover a standard theory of gravity and these new dofs are generically screened via a Vainshtein mechanism. This mechanism can be seen explicitly at work in braneworld models or models of massive gravity, such as DGP, NMG in three dimensions or Ghost-free massive gravity in four-dimensions. In all these models, the helicity-0 mode resembles in some limit a Galileon scalar field and can have an interesting phenomenology for the late-time acceleration of the Universe, be it self-acceleration or degravitation, if the strong coupling scale of the Galileon interactions is very low. For parametrically larger values of the strong coupling scale, the Galileon could on the other hand affect the very early Universe and in particular be a candidate for the  inflaton, leading to non-trivial non-gaussianities in the power spectrum of curvature fluctuations.

The screening of the helicity-0 mode at high energy implies that the departure from General Relativity on solar system scales are very much suppressed.  Nevertheless, this screening can yet have distinct signatures in cosmology and in particular for structure formation, \cite{Chow:2009fm}.
Another key signature of such models is the deviation from the standard Friedmann equation, that could distinguish these models from other scenarios of modified gravity or with additional dynamical degrees of freedom, see for instance Refs.~\cite{Afshordi:2008rd,Jain:2010ka}.

Another effect which can lead to specific signatures in models of massive gravity, is the emergence of a new type of coupling to matter of the form $\p_\mu \phi \p_\nu \phi T^{\mu\nu}$. Whilst this coupling is not relevant for spherically symmetric static sources as it vanishes in that case, it can be significant for relativistic sources. It has already been established by M.~Wyman in Refs.~\cite{Wyman:2011mp} and \cite{Sjors:2011iv}, that its effect in weak lensing could provide a smoking gun signature for this class of models.

Finally one peculiarity of Galileon models is the generic appearance of superluminal propagation, at least when the Vainshtein mechanism is at work, which may lead to the creation of Close-Timelike Curves (CTCs),  \cite{Adams:2006sv,Evslin:2011vh}. For instance if one of the dimensions were compact, CTCs may easily be created without the need of any additional exotic type of matter, however whenever one reaches a regime where CTCs may form, the Galileon  inevitably becomes infinitely strongly
coupled which leads to the breakdown of the effective field theory used to described the formation of CTCs in the first place. In addition, the formation of CTCs requires a background solution which is unstable with an arbitrarily fast decay time, similarly to a ghost-like instability. This suggests that an analogue of Hawking's chronology protection conjecture is also in effect for Galileons, \cite{Burrage:2011cr}.



\end{document}